%% file: iclr2025_conference.tex
\documentclass{article} 
\usepackage{iclr2025_conference,times}

\input{math_commands.tex}

\usepackage{hyperref}
\usepackage{url}
\usepackage{algorithm}
\usepackage{amssymb}
\usepackage{algpseudocode} 
\usepackage{chngcntr}
\counterwithout{algorithm}{section}
\usepackage{booktabs}
\usepackage{multirow}
\usepackage{appendix}
\usepackage{pdflscape}
\usepackage{graphicx}
\usepackage{amsmath,amssymb}
\usepackage{mathtools}
\usepackage{amsthm}
\usepackage{enumitem}
\usepackage{tabularx}
\usepackage{tabularx,array,booktabs}
\newcolumntype{P}[1]{>{\raggedright\arraybackslash}p{#1}}

\title{SparseEB-gMCR: A Generative Solver for Extreme Sparse Components with Application to Contamination Removal in GC-MS}

\author{
Yu-Tang Chang \quad Shih-Fang Chen \\
Department of Biomechatronics Engineering \\
National Taiwan University \\
\texttt{\{b05611038,sfchen\}@ntu.edu.tw}
}

\iclrfinalcopy 


\begin{document}

\maketitle
\begin{abstract}
Analytical chemistry instruments provide physically meaningful signals for elucidating analyte composition, and those with high mass or spectral resolution generate signals sparse enough for direct interpretation against chemical libraries.
Generative multivariate curve resolution (gMCR) models a sample as the linear superposition of a few components drawn from a learned pool, and its energy-based solver (EB-gMCR) recovers the pool and the component count without being told the count.
However, extreme sparsity in instruments such as GC-MS or ¹H-NMR leaves the component profiles themselves sparse, and a dense learnable profile cannot represent an exact zero.
To address this, a static support gate was introduced that applies the EB-select mechanism a second time, to the coordinates of each profile rather than to the components of each sample.
The result, SparseEB-gMCR, reparameterizes the gMCR pool rather than changing the model, and the sparsity that motivates the extension makes components easier to tell apart rather than harder.
On synthetic data, SparseEB-gMCR recovered the component count and reconstructed sparse mixtures as accurately as dense-component EB-gMCR, with the same graceful scaling in the number of components.
It was then applied to real GC-MS chromatograms for unsupervised contamination removal, where siloxane-related pollution signals were eliminated and compound identification became more reliable.
Removal reuses a pool learned from clean spectra alone, and rests on one requirement: that no combination of clean components can imitate the contamination, which is stronger than the two sets of components merely being different.
With this sparse extension, the EB-gMCR family becomes applicable to wider ranges of real-world chemical datasets, providing a general mathematical framework for signal unmixing and contamination elimination in analytical chemistry.
\end{abstract}

\section{Introduction}
Instruments in analytical chemistry elucidate chemical composition in analytes and provide valuable insight for material, biological, and food analysis \citep{He2006MaterialApproach, Aksenov2017GlobalChemAnalysis, Valdes2022FoodomicsChallenges}. 
Instruments that resolve individual chemical species often generate data in simpler and more interpretable forms, as seen in mass spectrometry (MS) and hydrogen-1 nuclear magnetic resonance (¹H-NMR). 
These techniques rely on sophisticated machinery and robust physical principles to ensure that each chemical component produces a unique signal pattern. 
In practice, the detected patterns are extremely sparse, enabling efficient and precise matching against pre-recorded chemical libraries. 
This sparse format also allows analysts to interpret results more directly. 
For instance, in a mass spectrum, the mass-to-charge (m/z) values of carbon dioxide appear at 12, 16, 28, and 44, corresponding to $\text{C}^+$, $\text{O}^+$, $\text{CO}^+$, and $\text{CO}_2^+$. 
In real data, isotopes of carbon and oxygen activate nearby m/z channels adjacent to these major responses. 
All values in the detected data carry physical meaning, which is why such instruments are regarded as the gold standard for elucidating unknown compounds. 

The collected data are not always sufficiently “clean” to allow straightforward identification of chemical composition.
Impurities in signal patterns can arise from many factors, including experimental design, device maintenance, or component aging.
More specifically, misconfigured chromatography, sample contamination, or neglected replacement of expired parts can all lead to noisy spectra.
Although sparsity generally improves clarity and efficiency in pattern matching—particularly when candidate patterns are already narrowed to a small set—it also becomes problematic when unexpected or uncertain patterns are present.
Noise may take the form of multiple overlapping signals or contamination of unknown origin, creating what is effectively an unknown signal mixture. 
Multivariate curve resolution (MCR) algorithms are well-suited to such mixtures \citep{deJuan2021MCR50Years} since signals from analytical instruments typically obey the Beer--Lambert law.
In some cases, external information such as chemical libraries or structural priors (e.g., peak-like patterns in 2D gas chromatography; \citealp{vanStokkum2009GlobalGCMS}) can improve decomposition of sparse chemical signals.
However, not all sparse data allow access to candidate sets, and in rare situations unknown patterns are absent from libraries altogether. 
External structures may also deviate from reality: for example, baseline correction or peak models proposed by \citet{vanStokkum2009GlobalGCMS} may not fit the collected data in some special cases.
For instance, certain contamination patterns may mimic real compounds while appearing across all retention times (RTs), and it is challenging to categorize such patterns into a peak or a baseline. 
For these reasons, when libraries are incomplete and external structures deviate from reality, a generalized MCR framework that remains effective under extreme sparsity and is capable of decomposing mixtures without external priors is still necessary.

MCR refers to a family of algorithms designed to decompose a set of $M$ mixed signals with $d$ features ($D \in \mathbb{R}^{M \times d}$ in Eq.~\ref{eq:1}) into base patterns (components, $S \in \mathbb{R}^{N \times d}$) and their corresponding intensities (concentrations, $C \in \mathbb{R}^{M \times N}$), with $\varepsilon$ the measurement noise.
In the classical setting of chemical signals, concentrations are non-negative, and the concentration matrix is sparse. 
In this study, the focus is further narrowed to conditions where the component matrix itself is also sparse.
MCR is generally formulated as a matrix factorization (MF) problem, with algorithms divided into iterative and non-iterative approaches.
Non-iterative MCR methods often require additional constraints, such as orthogonality, which may not reflect the structure of the data, particularly when the number of collected samples is insufficient to ensure uniqueness of Eq.~\ref{eq:1} \citep{Maeder1987EFA}. 
Iterative approaches are more widely used, including non-negative matrix factorization (NMF) and its extensions such as sparse NMF and Bayesian NMF \citep{Lee1999NMFParts, Hoyer2004NMFsparse, Schmidt2009BayesianNMF}. 
Among these, MCR–alternating least squares (ALS; \citet{Tauler1995SelectivityLocalRank}) is the most popular and has been broadly applied in chemical data analysis, especially spectroscopy \citep{Garrido2008MCRALSMonitoring, Felten2015MCRALS, Smith2019MCRALSRaman}. 
More recently, generative MCR (gMCR) and its energy-based solver (EB-gMCR; \citet{Chang2025EBgMCR}) re-modeled MCR as the process that generates a single sample rather than as a factorization of the data matrix.
Each sample activates a few components from a pool of candidates and is observed as their linear superposition plus noise.
Two properties follow that the experiments below rely on.
The number of components is determined by the data rather than supplied by the user, because selection is a variable of the search; and a recovered pool decodes a sample it has never seen, because the target of inference is the process and not the one matrix it was estimated from.

\begin{equation}
D = CS + \varepsilon
\label{eq:1}
\end{equation}

To adapt EB-gMCR for sparse component learning, SparseEB-gMCR applies the EB-select gating mechanism a second time, to the coordinates of each component profile rather than to the components of each sample, so that the zeros of a sparse profile are represented exactly.
This is a change of parameterization rather than a change of model.
In addition to benchmarking reconstruction performance against classical MCR methods under sparse conditions, this study applies SparseEB-gMCR to remove polluted patterns in a real-world gas chromatography--mass spectrometry (GC-MS) dataset where the contamination source is unknown. 
Although ground truth for quantitative evaluation of pollution removal is not available, quantitative MS analysis was performed to verify the effectiveness. 
This application illustrates how direct data generation modeling can advance real-world chemical data analysis. 
The results suggest that generative MCR frameworks can extend beyond synthetic validation to real-world problems, offering a new pathway for analyzing sparse chemical signals. 
Such capability may open further applications in mass spectrometry, spectroscopy, and other analytical platforms where sparsity is intrinsic.

\section{SparseEB-gMCR}
\label{sec:method}
\subsection{Preliminary}
\label{sec:2.1}
EB-gMCR replaces the factorization of Eq.~\ref{eq:1} with a model of how one sample is made \citep{Chang2025EBgMCR}.
A \emph{pool} is a set of unit-norm component profiles $s_1, \dots, s_N \in \mathbb{R}^d$, shared by every sample and learned from the data.
A sample, indexed by a draw $\omega$ of the generative process, activates a small subset of the pool and is observed as the linear superposition of the activated profiles plus measurement noise (Eq.~\ref{eq:2}).

\begin{equation}
x(\omega) \;=\; \sum_{i \in J(\omega)} c_i(\omega)\, s_i \;+\; \varepsilon(\omega),
\qquad
|J(\omega)| \le k
\label{eq:2}
\end{equation}

Here $J(\omega)$ is the sample's support, the set of components it contains, and $c_i(\omega)$ their concentrations.
The selection indicator $\delta_i(\omega) = \mathbf{1}[\,i \in J(\omega)\,]$ is what separates gMCR from classical MCR.
Classical MCR encodes absence through concentration magnitude: a component is absent when its concentration is near zero.
In gMCR, $\delta$ separates selection, which components participate, from intensity, how much each contributes, so component discovery is discrete rather than a shrinkage toward zero.
Only $x$ is observed; the pool, the supports, and the concentrations are latent and are what the solver infers.

Two counts are worth keeping apart, because a single symbol served for both in earlier presentations.
$N$ is the size of the candidate pool, a capacity set generously in advance and not a claim about the data.
$k$ caps how many components any one sample contains.
The number of pool slots the process actually uses is neither of these; it is the component count the solver is asked to discover, and the user supplies no value for it.

Two properties of this construction are what the present study relies on.
The first is self-determination of the component count, which is decided by the search rather than supplied, because selection is a variable of the optimization and not a fixed dimension.
The second is reuse: the recovered pool decodes a sample the solver has never seen, without re-fitting, because the target of inference is the generative process and not the one data matrix it was estimated from.
Both hold under conditions that \citet{Chang2025EBgMCR} states and analyzes: a linear detector response, a floor below which a component is not detectable, a limited dynamic range, few components per sample, and components distinct enough that no small group of them can imitate another.
This paper takes that construction as given and does not revisit the analysis.
What matters here is that mass spectra meet those conditions comfortably, and \S\ref{sec:2.2} explains why extreme sparsity helps rather than hinders on the last of them.

\subsection{A Support Gate for Extreme Sparse Profiles}
\label{sec:2.2}
The EB-select module, hereafter the selection gate, realizes the per-sample selection $\delta$.
For an observed sample $x$, a neural network $f_e$ evaluates selection energies $E(x) \in \mathbb{R}^{N \times 2}$, one energy $E_{j,1}$ for selecting component $j$ and one $E_{j,0}$ for not selecting it.
Learning a discrete selection by gradient is the obstacle, and the Gumbel--softmax reparameterization \citep{Jang2016GumbelSoftmax, Maddison2016Concrete} is the way around it: with $Z_{j,\ell} \sim \mathrm{Gumbel}(0,1)$ drawn independently, the gate is Eq.~\ref{eq:3}.

\begin{equation}
[f_e(x)]_j
\;=\;
\frac{\exp\bigl( ( Z_{j,1} - E_{j,1} ) / \tau \bigr)}
{\sum_{\ell \in \{0,1\}} \exp\bigl( ( Z_{j,\ell} - E_{j,\ell} ) / \tau \bigr)}
\label{eq:3}
\end{equation}

At temperature $\tau$ this is a smooth function of the energy gap $E_{j,0} - E_{j,1}$, so the selection probability is differentiable and gradients reach the energies.
Training uses a positive $\tau$ and perturbs the energies beforehand with additive Gaussian noise at a fixed scale, which explores the selection distribution without moving the parameters.
Evaluation uses a near-zero temperature, drops the perturbation, and selects component $j$ when its probability exceeds a fixed threshold.
Zeros in $\delta$ therefore mean non-selection, not small amplitudes.

Mass spectra call for a second kind of zero.
A profile $s_i$ is the fingerprint of the ionized fragments of one compound, and a zero coordinate is a fragment that the compound does not produce.
Such zeros belong to the profile, not to the sample: for a given component, the set of active m/z channels is fixed.
A dense learnable vector cannot express them, since gradient descent drives an unused coordinate toward zero without ever reaching it.

SparseEB-gMCR expresses them by applying the same gating mechanism to a different object.
A \emph{support gate} $f_{\mathrm{sg}}$ stores, for each pool slot $i$, a matrix of learnable energies $E^{\mathrm{sg}}_i \in \mathbb{R}^{d \times 2}$, one pair per coordinate, for that coordinate being active or inactive.
Eq.~\ref{eq:3} is applied to those energies coordinate by coordinate, giving a mask $m_i$ that is soft during training and thresholded to exact zeros and ones at evaluation.
The profile of slot $i$ is the masked dense vector, renormalized (Eq.~\ref{eq:4}).

\begin{equation}
s_i \;=\; \frac{m_i \odot u_i}{\| m_i \odot u_i \|} ,
\qquad
m_i \;=\; f_{\mathrm{sg}}\bigl( E^{\mathrm{sg}}_i \bigr)
\label{eq:4}
\end{equation}

with $u_i \in \mathbb{R}^d$ the learnable dense vector of slot $i$ and $\odot$ the coordinatewise product.
The two gates differ in what they read.
The selection gate reads the sample, because which components a sample contains varies from sample to sample.
The support gate reads nothing, because which coordinates a profile occupies does not; its energies are stored parameters rather than network outputs.
This is the sense in which the module is static.

The change is confined to how the pool is parameterized.
Eq.~\ref{eq:2} is untouched, $s_i$ remains a unit-norm vector in $\mathbb{R}^d$, and every downstream path---selection, concentration prediction, superposition---is unchanged.
SparseEB-gMCR is therefore a different parameterization of the same pool rather than a different model, which is why the construction of \S\ref{sec:2.1} continues to describe it.

Extreme sparsity also works in favor of the last condition listed in \S\ref{sec:2.1}, that no small group of components can imitate another.
Two compounds that fragment into different m/z channels produce profiles with little overlap, and profiles with little overlap are hard to confuse or to combine into one another.
Dense profiles, spread over every coordinate, are the harder case.
What does hurt is duplication, two slots drifting onto the same pattern, since they can then trade concentration freely; masking does nothing to prevent it.
That is why the distinctness term $\lambda_{\mathrm{amb}} \mathcal{R}(B)$ of Eq.~\ref{eq:5}, described in \S\ref{sec:2.3}, is carried over unchanged.

\subsection{Energy Optimization}
\label{sec:2.3}
The training loss is Eq.~\ref{eq:5}, with the support gate's energies entering where the selection gate's do.

\begin{equation}
\mathcal{L} \;=\; \| x - \hat{x} \|^2
\;+\; \lambda\, C(x)
\;+\; \lambda_{\mathrm{se}} \Bigl(
\underbrace{\| E \|_2^2}_{\textit{selection gate}}
\;+\;
\underbrace{\| E^{\mathrm{sg}} \|_2^2}_{\textit{support gate}}
\Bigr)
\;+\; \lambda_{\mathrm{amb}}\, \mathcal{R}(B),
\qquad
C(x) \;=\; \frac{1}{N} \sum_{i=1}^{N} [f_e(x)]_i
\label{eq:5}
\end{equation}

The reconstruction $\hat{x}$ is the superposition of the selected profiles weighted by the concentrations $f_c(x)$ of a concentration network that SparseEB-gMCR inherits unchanged, which is Eq.~\ref{eq:2} with the latent objects replaced by their learned counterparts.
The usage term $C(x)$ counts selections in soft form: as $\tau \to 0$ it tends to the number of selected components divided by $N$, so $\lambda\, C(x)$ is what pushes the solver toward reconstructing with fewer components.
The distinctness term $\mathcal{R}(B)$ penalizes similarity among the profiles in use in a minibatch $B$.
The energy regularizer $\lambda_{\mathrm{se}}$ removes the free common shift of each energy pair, and applies to $E^{\mathrm{sg}}$ for the same reason it applies to $E$, since only energy gaps affect a gate.
The single change from EB-gMCR is the added $\| E^{\mathrm{sg}} \|_2^2$ term, where $E^{\mathrm{sg}}$ stacks the per-slot energies.
Unless stated otherwise, $\tau$, the exploration scale, the gate threshold, $\lambda_{\mathrm{se}}$ and $\lambda_{\mathrm{amb}}$ take the values fixed across all EB-gMCR experiments \citep{Chang2025EBgMCR}.

The two gates settle differently, and no claim is made that they settle for the same reason.
The selection gate must track a moving input, since the energies it emits depend on the sample in front of it.
The support gate has no input to track: once the data have determined which coordinates a slot occupies, nothing sample-dependent moves those energies, and in practice its masks stop changing well before training ends.
What limits the learning of sparse profiles is therefore not the gate but the data, namely whether the collected samples exercise each component often enough, and in varied enough proportions, to pin down its support.

One quantity does require adjustment.
The support gate contributes $N d$ energy pairs against the selection gate's $N$, so at the profile dimensions used here its energies outweigh the usage term by three orders of magnitude, and a $\lambda$ tuned to the scale of the reconstruction error is overwhelmed before it can shape selection.
Raising $\lambda$ by roughly the factor $d$ restores the balance; in this study it was set to $10^{3}$ or above.
The practical cost is that the total energy no longer bottoms out sharply, so the automatic termination signal of EB-gMCR is less reliable here, and checkpoints are judged instead on reconstruction quality and component count.
Section~\ref{sec:exp} reports what this adjustment buys.

\subsection{GC-MS Contamination Removal as Two Processes over One Pool}
\label{sec:2.4}
Contamination in a GC-MS spectrum is not noise.
It is a second set of chemical profiles, obeying the same Beer--Lambert superposition as the analytes, and gMCR can represent it as such.
Let a clean process draw from a pool $S^{\mathrm{c}}$, and let a contamination process draw from a pool $S^{\mathrm{p}}$.
A polluted spectrum is then a sample of the combined process, whose pool is the union of the two and whose observation is their sum (Eq.~\ref{eq:6}).

\begin{equation}
x_{\mathrm{c}+\mathrm{p}}(\omega) \;=\; x_{\mathrm{c}}(\omega) \;+\; x_{\mathrm{p}}(\omega)
\label{eq:6}
\end{equation}

Separating that sum back into its two parts asks something of the union pool, and it is more than the two pools being different.
An earlier version of this construction required only that no profile appear in both, which is not enough: a contamination profile can be absent from the clean pool and still be closely imitated by a few clean profiles together, and when that happens the polluted spectrum admits two readings that fit equally well and nothing distinguishes them.
What is needed is that no small combination drawn from the two pools reproduces another profile of the union, which is the working form of the requirement that contamination be chemically distinguishable from the analytes.
For the column-fragment contamination studied here the requirement is plausible on chemical grounds, since siloxane fragments occupy m/z channels that the coffee volatiles largely do not, and the argument of \S\ref{sec:2.2} then applies: profiles on different channels are hard to confuse.

Under that requirement the removal follows without the contamination ever being observed, and it is the reuse property of \S\ref{sec:2.1} that makes it possible.
A solver is trained on clean spectra alone, learns $S^{\mathrm{c}}$, and is frozen.
Reuse is what licenses the next step: the frozen pool is a model of the clean process, not a fit to the clean matrix, so it can be applied to a spectrum drawn from a different process without being re-estimated.
Presented with a polluted spectrum, it can build reconstructions only from $S^{\mathrm{c}}$, and it cannot reach the contamination by rearranging clean profiles, because no combination of them imitates it.
What it rebuilds is the clean part, and the contamination is what remains in the residual (Eq.~\ref{eq:7}).

\begin{equation}
\hat{x}_{\mathrm{c}}(\omega) \;=\; \hat{x}\bigl( x_{\mathrm{c}+\mathrm{p}}(\omega) \bigr),
\qquad
\hat{x}_{\mathrm{p}}(\omega) \;=\; x_{\mathrm{c}+\mathrm{p}}(\omega) \;-\; \hat{x}_{\mathrm{c}}(\omega)
\label{eq:7}
\end{equation}

Here $\hat{x}(\cdot)$ is the reconstruction produced by the solver frozen on $S^{\mathrm{c}}$, so the cleaned spectrum is what that solver rebuilds and the estimated contamination is what it leaves behind.

Two features of gMCR carry this argument, and a factorization has neither.
The first is the reuse property just used, which a factorization cannot offer: a factorization computed on the clean matrix describes that matrix and says nothing about a new sample.
The second is that selection is explicit.
Over the union pool, the statement that a sample carries no contamination is the statement that $\delta_i = 0$ for every slot whose profile belongs to $S^{\mathrm{p}}$, a well-defined event.
In a factorization the same statement is a claim about coefficient magnitudes with no threshold to appeal to, so there is no point at which a concentration column may be declared inactive.

The construction generalizes: given any two of the three processes, clean, contamination, and polluted, the third is determined, so the roles of $S^{\mathrm{c}}$ and $S^{\mathrm{p}}$ may be exchanged wherever the contamination is characterized and the analytes are not.
Different processes can also be trained separately and combined afterwards.
What every such case needs is the distinguishability requirement above, and that requirement concerns profiles that are not all observed.
It can be argued chemically, as it is here, and it can be inspected on a returned solution, but it cannot be verified in advance.
\section{Evaluation on Synthetic Dataset with Extreme Sparse Components}
\label{sec:exp}
\subsection{Generation of the Sparse-Component Synthetic Data}
\label{sec:3.1}
SparseEB-gMCR was primarily evaluated on synthetic datasets, since real-world chemical data rarely provide cleanly separated mixtures with known concentrations and components. 
In practice, pure and uncontaminated spectra for MS or ¹H-NMR exist only in commercial libraries or those of the National Institute of Standards and Technology (NIST) \citep{Heller1999HistoryNISTEPAHIH}. 
For convenience and full controllability, SparseEB-gMCR was tested on synthetic datasets with adjustable sparsity ratios. 
When the ratio is set to 0.9, for example, 90\% of the indices in each sampled component are zero, and the remaining non-zero entries are scaled to form a unit vector, simulating extreme sparsity. 
This ratio represents the average sparsity across all components, rather than the exact value for any individual one.

In all experiments, the sparsity ratio was fixed at 0.95.
To simulate the characteristics of MS spectra targeted in this study, datasets were generated following the same procedure as in EB-gMCR \citep{Chang2025EBgMCR}, with minor modifications to the concentration range.
Concentrations were sampled from 10 to 1,000 and converted to integers after adding Gaussian noise to the mixed data.
SparseEB-gMCR was benchmarked against NMF, sparse-NMF, Bayes-NMF, independent component analysis (ICA), and MCR-ALS using datasets of $4N$ and $8N$ sizes under two noise levels (20 dB and 30 dB).
Following the convention of \citet{Chang2025EBgMCR}, a dataset size of $4N$ or $8N$ means four or eight samples per true component; the candidate pool the solver is given is a separate and much larger quantity.
The number of components a method reports is its estimated component count (EC).
The overall evaluation workflow followed that of EB-gMCR \citep{Chang2025EBgMCR}. 

\subsection{Component-Count Estimation and Reconstruction on Sparse-Component Data}
\begin{figure}[t]
\centering
\includegraphics[width=1\linewidth]{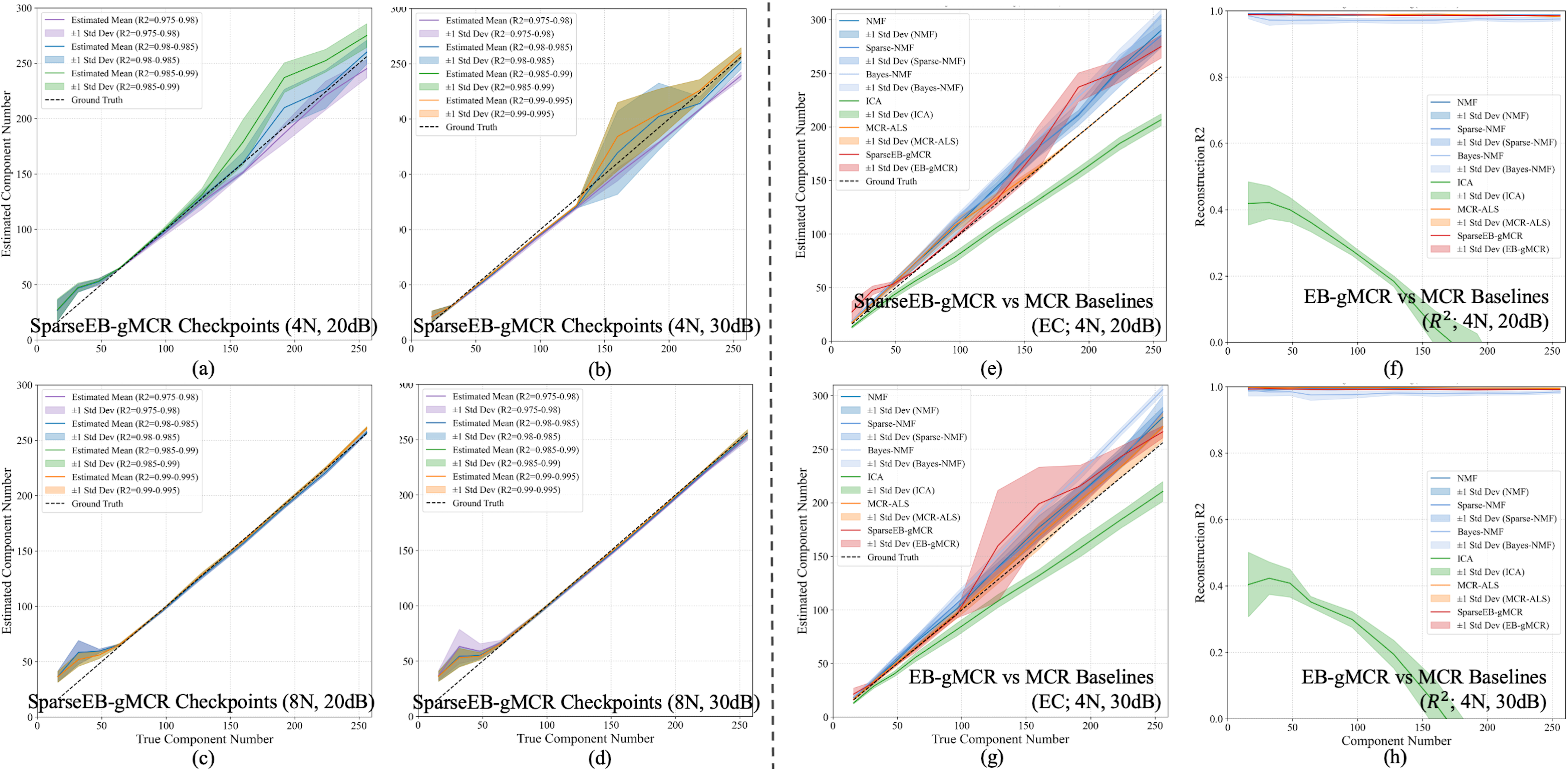}
\caption{Synthetic benchmarks. (a–d) SparseEB-gMCR checkpoints: estimated vs. true component number (dashed black line); mean (solid) and ±1 SD (shaded) over 5 replicates; colors denote $R^2$ checkpoint bands. Panels: (a) $4N$, 20 dB; (b) $4N$, 30 dB; (c) $8N$, 20 dB; (d) $8N$, 30 dB. (e, g) SparseEB-gMCR vs. baselines: estimated vs. true components at $4N$ under 20 dB and 30 dB. (f, h) Reconstruction $R^2$ at each method’s estimated component count (EC) for the same settings.}
\label{fig:1}
\end{figure}

In the synthetic dataset, the SparseEB-gMCR solvers recovered the component count as accurately as EB-gMCR applied to dense components (Fig.~\ref{fig:1}).
The solvers reconstructed sparse signals with high accuracy, achieving $R^2$ values exceeding 0.99—a level of performance not observed in the dense version of EB-gMCR.
It should be noted that the evaluation considered both zero and non-zero elements; therefore, lower $R^2$ values may occur when the solver fails to capture the extreme sparsity and wide intensity range inherent in the synthesized data. 
Similar to its dense counterpart, SparseEB-gMCR maintained strong performance in reconstructing data using the fewest components and demonstrated graceful scalability as the number of components increased (Fig.~\ref{fig:1}a). 
The introduction of a support gate to handle extreme sparsity did not impair recovery. 
The primary difference from the dense version lies in the tuning of the usage penalty $\lambda$.
In SparseEB-gMCR, this weight was raised in proportion to the dimensionality of the learned profiles, for the reason given in \S\ref{sec:2.3}. 
This simple adjustment proved sufficient for stabilizing energy optimization and enabling effective handling of extreme sparsity. 

The bar is lower here than in the dense setting, however.
When benchmarked against other MF-based MCR methods, the reconstruction performance of SparseEB-gMCR did not differ as markedly as that observed in the dense-profile experiments with EB-gMCR.
Achieving $R^2$ values above 0.99 was not particularly challenging for traditional MCR approaches in this sparse setting.
The high proportion of zero entries in the data also had limited impact on reconstruction accuracy when the component number was small. 
However, closer inspection of the reconstructed values revealed that several zero entries in the original sparse data were reconstructed as small non-zero values (e.g., 0.001). 
In contrast, SparseEB-gMCR enforced exact zeros through its support gate, producing a more precise masking effect.
Although this difference has minimal influence on most real-world chemical analyses, the ability to enforce true zero activation can be important for datasets where strict sparsity is physically meaningful. 
Following the validation workflow of EB-gMCR, SparseEB-gMCR reconstructed the mixtures to the accuracy the noise permitted while recovering the correct number of components under extreme sparsity. 

\section{Contamination Removal in GC-MS Chromatogram}
\subsection{Collected GC-MS Spectra with Unknown Contamination}
A total of 456 GC-MS spectra were collected using solid-phase microextraction (SPME) from coffee odor samples, including two major types: fragrance (from dry ground coffee powder) and aroma (from brewed coffee). 
All spectra were obtained from distinct coffee samples, ensuring that no sample was measured more than once.
The chromatographic analyses were conducted using an Agilent 7890A–5975C GC/MSD system equipped with a DB-Wax column (Agilent Technologies Inc., Santa Clara, CA, USA). 
The complex odor mixtures were separated over a 55-minute run, with the oven temperature programmed to increase from 30 °C to 250 °C.
The mass spectrometer was configured to detect ions over an m/z range of 12–501. 
Each GC-MS chromatogram therefore forms a sparse matrix of size $3375 \times 490$.
After manual inspection, 48 chromatograms were identified as polluted, primarily due to glass-like fragments detaching from the DB-Wax column. Further details of contamination identification are provided in Appendix~\ref{app:a}.

\subsection{SparseEB-gMCR for GC-MS Chromatogram Modeling}
\begin{figure}[t]
\centering
\includegraphics[width=0.8\linewidth]{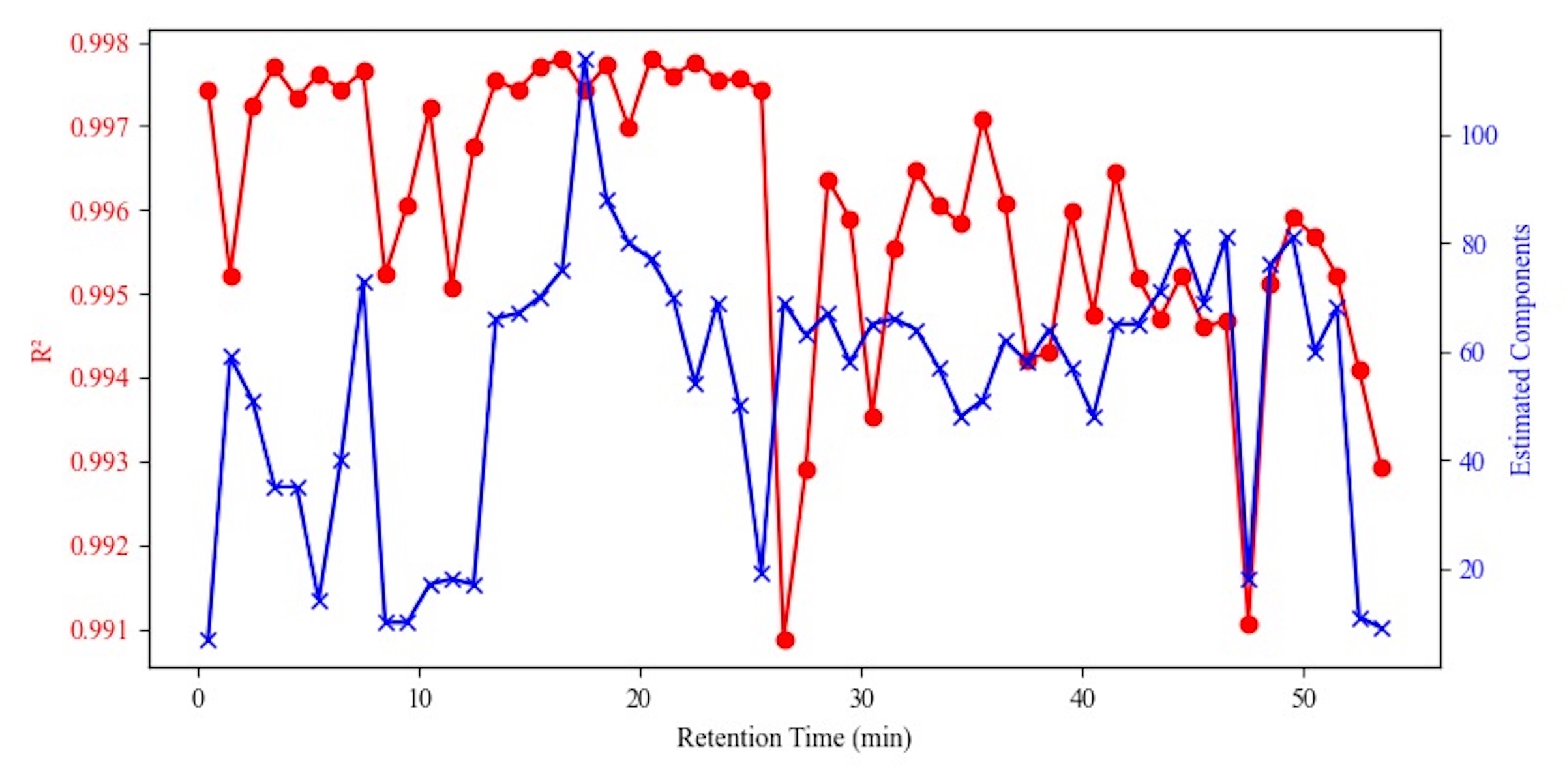}
\caption{Reconstruction $R^2$ and estimated components (EC) of SparseEB-gMCR on 408 clean GC-MS chromatograms.}
\label{fig:2}
\end{figure}

In GC-MS chromatograms, volatile organic compounds (VOCs) within odor samples are separated through chromatographic retention. 
Mass patterns detected at distant RTs therefore correspond to distinct VOCs.
Dividing the RT axis into smaller intervals (e.g., 0–1 min, 1–2 min, etc.) provides a practical strategy to reduce computational cost, and it also keeps the number of co-eluting compounds small within each interval, which is what the distinguishability requirement of \S\ref{sec:2.4} needs in order to hold locally.
Since clean GC-MS chromatograms represent the majority of the dataset, the data generation process of 408 clean chromatograms was modeled using SparseEB-gMCR in this study. 
These trained solvers were subsequently applied to contaminated chromatograms to eliminate pollution patterns according to Eq.~\ref{eq:7}.
All SparseEB-gMCR solvers were initialized with a candidate pool of $N = 1{,}024$ slots and trained for up to 100,000 iterations to ensure convergence. 
Although only clean samples were used for training, this approach was sufficient for modeling the major mass patterns, as discussed further in \S\ref{sec:4.4}.
The best checkpoints were selected based on reconstruction accuracy, with all solvers achieving $R^2$ values between 0.991 and 0.998 for clean chromatograms (Fig.~\ref{fig:2}).
The number of estimated components ranged from 7 to 114.

Our earlier analyses of this dataset indicated that most VOC peaks observed in total ion chromatograms (TICs) appear within RTs of 5–35 minutes.
In practice, the number of compounds eluted within a given RT interval is small. 
Even VOCs with similar molecular structures typically elute within close but distinct RTs, making it unrealistic for more than 10 compounds to co-elute within the same minute.
Therefore, cases where SparseEB-gMCR estimated 20 or more components likely indicate overfitting to background noise, the expected outcome when the usage penalty is too weak to suppress the last few slots at the sample size available. 
Interestingly, this overfitting is not necessarily detrimental.
In SparseEB-gMCR, it reflects the solver’s capacity to reconstruct nearly all spectral variations within clean chromatograms. 
When such a solver is applied to polluted chromatograms, which are samples of the combined process of Eq.~\ref{eq:6} and therefore cannot be represented by $S^{\mathrm{c}}$ alone, it is expected to reconstruct the clean portion of the signal while failing to reproduce the unknown pollution patterns.

\subsection{Mass Patterns and VOC Analysis of Polluted GC-MS Chromatogram}
\begin{figure}[t]
\centering
\includegraphics[width=1.\linewidth]{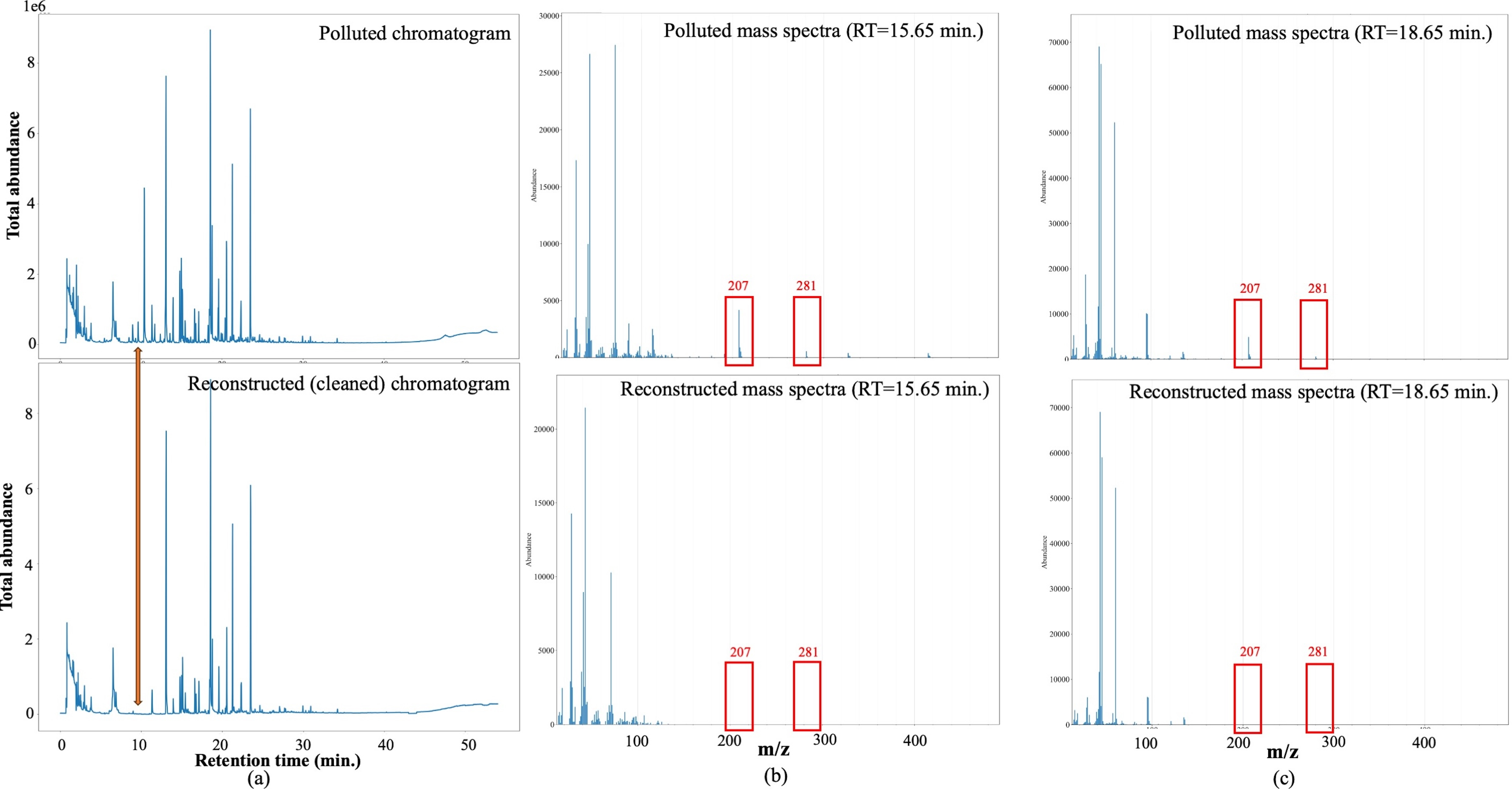}
\caption{Comparison of the polluted and reconstructed (cleaned) chromatogram of a Brazil coffee powder odor sample. (a) Total ion chromatogram (TIC); (b) Mass spectrum at RT = 15.65 min; (c) Mass spectrum at RT = 18.65 min.}
\label{fig:3}
\end{figure}

Instead of quantitatively evaluating reconstruction performance on polluted samples, this section focuses on how SparseEB-gMCR improves the quality of mass patterns and thereby enhances chemical interpretability. 
A GC-MS chromatogram obtained from dried Brazil specialty coffee powder was randomly selected from the polluted set for illustration (Fig.~\ref{fig:3}).
It should be noted that the reconstructed mass spectra produced by SparseEB-gMCR do not rely on any RT-related features such as peak modeling. 
The reconstructed spectra are generated solely from the components stored in the optimized solver. 
The observed peak shapes in the TIC arise because the concentration predictor accurately estimates intensity from the observed polluted mass patterns at each time point. 
By comparing the original polluted chromatogram with the reconstructed one (Fig.~\ref{fig:3}a), most TIC peaks are reproduced correctly, except for those near a retention time of approximately 10 minutes.
Given the relatively lower $R^2$ ($\sim$0.995) and smaller number of components used by the SparseEB-gMCR solver in this RT region, the missing peak may result from limited reconstruction capacity in that segment.
However, an alternative explanation is that the corresponding VOC mass pattern does not appear in any of the clean samples used for training. This possibility is discussed in more detail in \S\ref{sec:4.4}. 

To further verify the differences in mass spectra reconstructed by SparseEB-gMCR, two representative peaks were examined (RT = 15.65 min and RT = 18.65 min; Fig.~\ref{fig:3}b,~\ref{fig:3}c).
Both spectra were extracted at the centroid of each peak to ensure that the analyzed mass patterns are chemically meaningful. 
These spectra were produced by SparseEB-gMCR solvers trained on different RT intervals (15–16 min and 18–19 min, respectively), meaning that each was generated by a distinct solver. 
The peak at 15.65 min corresponds to a normal VOC peak, whereas the one at 18.65 min represents the highest peak in the chromatogram. 
As described earlier, the contamination originates from fragments detached from the GC column. 
A previous report \citep{English2022SiloxaneGhostPeaks} comprehensively described this type of contamination, identifying m/z 207 and 281 as characteristic of siloxane-related signals. 
Although direct evidence confirming that the same siloxane fragments from the DB-Wax column produce these m/z channels is unavailable, this analysis focuses on how their intensities change after cleaning by SparseEB-gMCR. 
In the polluted mass spectra at both RT = 15.65 min and RT = 18.65 min, notable abundances are observed at m/z 207 ($\sim$4000) and m/z 281 ($\sim$400).
These signals completely disappear in the reconstructed (cleaned) spectra.

The reconstructed spectra were further examined using the MassBank compound identification algorithm (Table~\ref{tab:1}; \citealp{Horai2010MassBank}).
Considering that the chromatograms were obtained from coffee odors, the compounds identified in the cleaned spectrum at RT = 15.65 min correspond more closely to VOCs commonly reported in coffee \citep{Buettner2017SpringerHandbookOdor, Caporaso2018SingleBeanVolatiles}. 
This improvement is even more evident in the stronger peak at 18.65 min, where the order of identified compounds remains unchanged but the confidence scores increase substantially after removal of siloxane-related pollution.
Although direct chemical verification was not possible due to the instability and rapid intensity decay of coffee odor samples, the collective evidence supports that SparseEB-gMCR effectively eliminates unknown mass contamination.
While only one chromatogram is shown, the same qualitative change, loss of m/z 207 and 281 together with higher MassBank confidence, was observed for the other polluted chromatograms examined. 
The case shown is therefore representative rather than selected, and it illustrates how SparseEB-gMCR provides a mathematically grounded and chemically consistent approach to removing unknown contamination.

\begin{table}[htbp]
\centering
\caption{Compound identification from MassBank for chromatogram peaks at RT = 15.65 min and RT = 18.65 min.}
\label{tab:1}
\renewcommand{\arraystretch}{1.05}
\setlength{\tabcolsep}{2pt}
\scriptsize
\begin{tabular}{c|P{3.8cm}|l|r||P{3.8cm}|l|r|}
\hline
\textbf{Order} &
\multicolumn{3}{c||}{\textbf{Original mass spectra (RT = 15.65 min)}} &
\multicolumn{3}{c|}{\textbf{Cleaned mass spectra (RT = 15.65 min)}} \\
\cline{2-7}
 & \textbf{Compound} & \textbf{Formula} & \textbf{Score} 
 & \textbf{Compound} & \textbf{Formula} & \textbf{Score} \\
\hline
1  & TRIACONTANE & C$_{30}$H$_{62}$ & 0.5773 & 2-METHYLTETRAHYDROFURAN & C$_5$H$_{10}$O & 0.8235 \\
2  & 2-METHYLTETRAHYDROFURAN & C$_5$H$_{10}$O & 0.5560 & ALLYL BUTYRATE & C$_7$H$_{12}$O$_2$ & 0.7825 \\
3  & 2-METHYLTETRAHYDROFURAN & C$_5$H$_{10}$O & 0.5285 & 2,3-DIMETHYLBUTANOL & C$_6$H$_{14}$O & 0.7690 \\
4  & 3-CARBOETHOXY-6-ISOPROPYL-1-OXA-3,3A,4,5,6,7-HEXAHYDROAZULENE-2(1H),8(8AH)-DIONE
   & C$_{15}$H$_{22}$O$_5$ & 0.5240 & 2-METHYLPENTANE & C$_6$H$_{14}$ & 0.7687 \\
5  & 1,1-DIALLYLHYDRAZINE & C$_6$H$_{12}$N$_2$ & 0.5187 & ALLYL ISOBUTYRATE & C$_7$H$_{12}$O$_2$ & 0.7683 \\
6  & ALLYL BUTYRATE & C$_7$H$_{12}$O$_2$ & 0.5148 & (2S,3R)-(-)-2,3-Epoxyoctanal & C$_8$H$_{14}$O$_2$ & 0.7587 \\
7  & VINYL BUTYRATE & C$_6$H$_{10}$O$_2$ & 0.5127 & METHALLYL BUTYRATE & C$_8$H$_{14}$O$_2$ & 0.7565 \\
8  & 2-BROMOPENTANE & C$_5$H$_{11}$Br & 0.5093 & 2-BROMOPENTANE & C$_5$H$_{11}$Br & 0.7521 \\
9  & TRIDECANE & C$_{13}$H$_{28}$ & 0.5091 & VINYL BUTYRATE & C$_6$H$_{10}$O$_2$ & 0.7511 \\
10 & TETRAHYDROFURFURYL ALCOHOL & C$_5$H$_{10}$O$_2$ & 0.5089 & TETRAHYDROFURFURYL ALCOHOL & C$_5$H$_{10}$O$_2$ & 0.7458 \\
\hline
\textbf{Order} &
\multicolumn{3}{c||}{\textbf{Original mass spectra (RT = 18.65 min)}} &
\multicolumn{3}{c|}{\textbf{Cleaned mass spectra (RT = 18.65 min)}} \\
\cline{2-7}
 & \textbf{Compound} & \textbf{Formula} & \textbf{Score} 
 & \textbf{Compound} & \textbf{Formula} & \textbf{Score} \\
\hline
1  & ACETIC ACID & C$_2$H$_4$O$_2$ & 0.6565 & ACETIC ACID & C$_2$H$_4$O$_2$ & 0.8285 \\
2  & MALONIC ACID & C$_3$H$_4$O$_4$ & 0.5989 & MALONIC ACID & C$_3$H$_4$O$_4$ & 0.7341 \\
3  & 1,1-DIMETHYLHYDRAZINE & C$_2$H$_8$N$_2$ & 0.5770 & 1,1-DIMETHYLHYDRAZINE & C$_2$H$_8$N$_2$ & 0.7224 \\
4  & METHOXYACETIC ACID & C$_3$H$_6$O$_3$ & 0.5669 & MALONIC ACID & C$_3$H$_4$O$_4$ & 0.7042 \\
5  & 3-METHYLBUTANOIC ACID & C$_5$H$_{10}$O$_2$ & 0.5653 & METHOXYACETIC ACID & C$_3$H$_6$O$_3$ & 0.7033 \\
6  & MALONIC ACID & C$_3$H$_4$O$_4$ & 0.5644 & METHOXYACETIC ACID & C$_3$H$_6$O$_3$ & 0.6922 \\
7  & ETHYLHYDRAZINE & C$_2$H$_8$N$_2$ & 0.5575 & ETHYLHYDRAZINE & C$_2$H$_8$N$_2$ & 0.6908 \\
8  & METHOXYACETIC ACID & C$_3$H$_6$O$_3$ & 0.5556 & 3-METHYLBUTANOIC ACID & C$_5$H$_{10}$O$_2$ & 0.6900 \\
9  & ISOVALERIC ACID & C$_5$H$_{10}$O$_2$ & 0.5515 & ISOVALERIC ACID & C$_5$H$_{10}$O$_2$ & 0.6805 \\
10 & ETHYL METHYL ETHER & C$_3$H$_8$O & 0.5439 & ETHYL METHYL ETHER & C$_3$H$_8$O & 0.6795 \\
\hline
\end{tabular}
\end{table}

\subsection{Limitation and Concern}
\label{sec:4.4}
Although both quantitative reconstruction results and chemical analyses demonstrate the effectiveness of SparseEB-gMCR, several concerns remain regarding its underlying construction inherited from EB-gMCR.
The modeled data generation process can nearly perfectly reproduce mixed signals, in both synthetic and real chemical datasets, only if the underlying patterns are present in the training set. 
In the coffee odor dataset used here, the polluted and clean sample sets do not overlap, and the VOC composition of coffee inherently varies due to agricultural and processing factors such as cultivation, fermentation, and roasting.
Consequently, some VOC patterns appearing in polluted samples may never have been observed in the clean set. 
This limitation does not indicate a chemically meaningless pattern but reflects a general mathematical constraint inherent to data-driven generative modeling: EB-gMCR can only reconstruct or eliminate patterns that lie in the span of its learned pool.
In other words, the effect arises purely from which profiles the two pools $S^{\mathrm{c}}$ and $S^{\mathrm{p}}$ of Eq.~\ref{eq:6} contain, rather than from any bias toward specific compounds. 
Hence, both the assumed data-generation formulation and the design of the dataset critically determine the number and diversity of component profiles the solver can model.

The success of contamination elimination thus depends on two factors: the strength of the EB-gMCR solver in signal unmixing and the correctness of data integration in defining the two pools $S^{\mathrm{c}}$ and $S^{\mathrm{p}}$. 
Regardless of solver capability, patterns absent from the training data cannot be generated. 
In the context of GC-MS analysis, taking the entire mass spectral library as the clean pool $S^{\mathrm{c}}$ and device-related background signals as the contamination pool $S^{\mathrm{p}}$ could provide a more universal SparseEB-gMCR model applicable across instruments, as discussed by \citet{Marty2015BayesianDeconvolution}.
Such a pool would have to be checked for distinguishability rather than assumed to have it, since a library large enough to cover many instruments is also large enough to contain near-duplicate patterns.
Such an integration would further enhance the potential of gMCR to perform unsupervised contamination elimination in diverse GC-MS applications.

\section{Conclusion}
EB-gMCR recovers the components and the component count of dense component and spectral datasets by modeling how a sample is generated rather than factorizing the data matrix. 
In this study, the framework was extended to a sparse variant, SparseEB-gMCR, by introducing a static support gate that applies the EB-select mechanism to each coordinate of each component profile.
The gate imposes a hard, learned mask on the dense profile of every pool slot, which is then renormalized, so the pool is reparameterized while the generative model of EB-gMCR is left intact. 
Following the energy optimization principle of EB-gMCR, the usage penalty was raised in proportion to the profile dimension, so that it is not overwhelmed by the energies the support gate adds. 
The proposed SparseEB-gMCR was validated on synthetic datasets and further applied to real GC-MS chromatograms of coffee odor samples for contamination removal. 
The results demonstrated that SparseEB-gMCR retains the reconstruction quality and the self-determined component count of EB-gMCR while extending its adaptability to sparse and irregular chemical data. 
By complementing the EB-gMCR framework with this sparse formulation, the gMCR family broadens its applicability to complex and diverse real-world mixture problems, providing a general mathematical tool for signal unmixing and contamination elimination in analytical chemistry. 

\subsubsection*{Author Contributions}
All conceptual development, experimental design, and implementation were carried out by the first author. The second author is listed per lab authorship policy.

\nocite{*}
\bibliographystyle{iclr2025_conference}
\bibliography{iclr2025_conference}

\clearpage
\appendix
\section*{Appendix}
\addcontentsline{toc}{section}{Appendix}  

\renewcommand{\thesection}{A}
\section{Contamination Identification in GC-MS Chromatograms}
\label{app:a}
The GC-MS chromatograms were manually collected by the first author as part of a long-term study on olfactory cognition and chemical perception in specialty coffee.
Data collection was conducted over several years using a shared GC-MS facility at the College of Bioresources and Agriculture, National Taiwan University. 
During this period, a contamination incident occurred in which the GC column was damaged by samples with improper derivatization processing.

The contamination was not immediately detected because the column fragment signal did not appear as a visible peak in the TIC. 
Instead, it manifested as a low-intensity, background-like signal. As a result, automatic compound identification algorithms—typically focused on large peaks of VOCs—continued to yield seemingly normal results. 
Several months later, irregular mass spectrometric responses were observed, and the column was subsequently replaced.

In coffee aroma analysis, small peaks often correspond to characteristic sensory descriptors (e.g., floral or citrus notes) rather than major VOCs defining the general “coffee” odor. 
Consequently, background signals from column fragments become non-negligible and can interfere with these fine features. 
Preliminary inspection indicated that, even with knowledge of the contamination source (e.g., glass-like fragments), separating the polluted signals by chemical heuristics such as the m/z channel of $\text{Si}^+$ was highly challenging. 
The major difficulty arose from the continuous and non-sparse nature of the contamination background, unlike the distinct sparse patterns stored in MS libraries. 
Although a threshold-based filter could theoretically reduce background responses, determining an appropriate threshold becomes unreliable when contamination overlaps with meaningful signals.

This challenge motivated the development of SparseEB-gMCR, which suppresses unknown pollution patterns through the two-process construction of Eqs.~\ref{eq:6}--\ref{eq:7}, even when the exact contamination profile is unavailable.

\end{document}

%% file: math_commands.tex
\usepackage{amsmath,amsfonts,bm}

\def\eqref#1{equation~\ref{#1}}

\def\1{\bm{1}}

\DeclareMathAlphabet{\mathsfit}{\encodingdefault}{\sfdefault}{m}{sl}
\SetMathAlphabet{\mathsfit}{bold}{\encodingdefault}{\sfdefault}{bx}{n}

